\documentclass{elsart}

\usepackage{graphicx}

\usepackage{amssymb}
\usepackage{amsmath}

\begin{document}

\setlength{\pdfpageheight}{297mm}
\setlength{\pdfpagewidth}{210mm}
\newlength{\adlt}
\setlength{\adlt}{20pt}

\begin{frontmatter}



\title{Nab: Measurement Principles, Apparatus and Uncertainties}


\author[uva]{Dinko Po\v{c}ani\'c\corauthref{cor1}},
\author[asu]{R. Alarcon},
\author[uva]{L.P. Alonzi},
\author[uva]{S. Bae{\ss}ler},
\author[asu]{S. Balascuta},
\author[ornl]{J.D. Bowman},
\author[uva]{M.A. Bychkov},
\author[suss]{J. Byrne}, 
\author[unh]{J.R. Calarco},
\author[ornl]{V. Cianciolo},
\author[uky]{C. Crawford},
\author[uva]{E. Frle\v{z}},
\author[manitoba]{M.T. Gericke},
\author[utk]{G.L. Greene},
\author[utk]{R.K. Grzywacz},
\author[usc]{ V. Gudkov},
\author[unh]{F.W. Hersman},
\author[lanl]{A. Klein},
\author[winnipeg]{J. Martin},
\author[manitoba]{S.A. Page},
\author[uva]{A. Palladino},
\author[ornl]{S.I. Penttil\"a},
\author[ornl]{K.P. Rykaczewski},
\author[lanl]{W.S. Wilburn},
\author[ncst]{A.R. Young},
\author[ornl]{G.R.~Young}
\collaboration{The Nab collaboration}
 
\corauth[cor1]{Corresponding author: {\tt pocanic@virginia.edu}}
\address[uva]{Department of Physics, University of Virginia,
  Charlottesville, VA 22904, USA}
\address[asu]{Department of Physics, Arizona State University, Tempe, AZ
   85287-1504, USA}  
\address[uky]{Department of Physics and Astronomy, University of Kentucky,
  Lexington, Kentucky 40506, USA}
\address[manitoba]{Department of Physics and Astronomy, 
   University of Manitoba, Winnipeg, Manitoba, R3T 2N2 Canada}
\address[lanl]{Los Alamos National Laboratory,  Los Alamos, NM 87545, USA}
\address[unh]{Department of Physics, University of New Hampshire,
  Durham, NH 03824, USA}
\address[ncst]{Department of Physics, North Carolina State University, Raleigh,
   NC 27695-8202, USA}
\address[ornl]{Physics Division, Oak Ridge National Laboratory, Oak
  Ridge, TN 37831, USA}
\address[usc]{Department of Physics and Astronomy, University of South
   Carolina, Columbia, SC 29208, USA} 
\address[suss]{Department of Physics and Astronomy, University of
  Sussex, Brighton, BN19RH, UK}
\address[utk]{Department of Physics and Astronomy,  University of Tennessee,
    Knoxville, TN 37996-1200, USA}
\address[winnipeg]{Department of Physics, University of Winnipeg,
  Winnipeg, Manitoba, R3B 2E9 Canada}

\newpage

\begin{abstract}  

The Nab collaboration will perform a precise measurement of $a$, the
electron-neutrino correlation parameter, and $b$, the Fierz
interference term in neutron beta decay, in the Fundamental Neutron
Physics Beamline at the SNS, using a novel electric/magnetic field
spectrometer and detector design.  The experiment is aiming at the
$10^{-3}$ accuracy level in $\Delta a/a$, and will provide an
independent measurement of $\lambda = G_A/G_V$, the ratio of
axial-vector to vector coupling constants of the nucleon.  Nab also
plans to perform the first ever measurement of $b$ in neutron decay,
which will provide an independent limit on the tensor weak coupling.

\end{abstract}

\begin{keyword}
neutron beta decay \sep correlations \sep precision measurement
\PACS 13.30.-Ce \sep 14.20.Dh \sep 23.40.-s \sep 24.80.-y
\end{keyword}
\end{frontmatter}

\section{Motivation}
\label{sec:motiv}

Neutron beta decay provides one of the most sensitive means for
exploring details and limits of our understanding of the weak
interaction.  Thanks to its highly precise the\-o\-re\-ti\-cal
description \cite{Cza04}, neutron decay is sensitive to contributions
from processes not included in the standard model (SM) of particles
and interactions (for comprehensive reviews see
Refs.~\cite{Her01,Sev06,Ram08}).  Neglecting recoil, radiative and
loop corrections, the differential decay rate for unpolarized neutrons
is given by parameters $a$ and $b$: $dw \propto 1 +
a\beta_{\text{e}}\cos\theta_{\text{e}\nu} +
b(m_{\text{e}}/E_{\text{e}})$, where
$\beta_{\text{e}}=p_{\text{e}}/E_{\text{e}}$, $p_{\text{e}}$,
$E_{\text{e}}$ and $\theta_{e\nu}$ are the electron momentum, energy,
and e--$\nu$ opening angle, respectively \cite{Jac57}.  The
e--$\nu_{\rm e}$ correlation parameter $a$, and the asymmetry
parameters with respect to the neutron spin: $A$ (beta), $B$
(neutrino), and $C$ (proton; $C \propto A+B$ in leading order) possess
complementary dependencies on the ratio of Fermi constants
$\lambda=G_A/G_V$, as well as on operators that depart from the
$(V-A)\otimes(V-A)$ form of the SM charged current (CC) weak
interaction.  Additionally, $b$, the Fierz interference term, offers
an independent test of scalar and tensor admixtures arising in broad
classes of L-R mixing SUSY extensions.  Thus precise measurements of
neutron decay parameters offer the distinct advantage of
overconstrained independent checks of the SM predictions, as well as
the potential for indicating or ruling out certain types of extensions
to the SM $(V-A)\otimes(V-A)$ form \cite{Her01,Sev06,Ram08,Gud06}.
Hence, a set of appropriately precise measurements of the neutron
decay parameters $a$, $b$, $A$, and $B$ will have considerably greater
physics implications than the erstwhile predominant experimental focus
on $A$, i.e., $\lambda$.  At a minimum, such a data set combined with
new measurements of the neutron lifetime, $\rm \tau_n$, will enable a
definitive resolution of the persistent discrepancies in $\lambda$ and
Cabibbo--Kobayashi-Maskawa (CKM) matrix element $V_{ud}$ \cite{PDG06}.

The Nab collaboration \cite{nab} has undertaken to carry out precise
measurements of $a$, the e--$\nu_{\rm e}$ correlation parameter, and
$b$, the so far unmeasured Fierz interference term, in neutron decay.
Goal accuracies are $\Delta a/a \simeq 10^{-3}$ and $\Delta b \simeq
10^{-3}$.  A novel $4\pi$ field-expansion spectrometer based on ideas
outlined in Ref.~\cite{Bow05} will be used in the Fundamental Neutron
Physics Beamline (FnPB) at the Spallation Neutron Source (SNS) at Oak
Ridge, Tennessee.

The Nab experiment constitutes the first phase of a program of
measurements that will continue with second-generation measurements of
spin correlations in neutron decay.  The next experiment, named
`abBA', will measure parameters $A$ and $B$ in addition to $a$ and
$b$.  In addition, the proton asymmetry $C$ will be measured with the
same apparatus.  Together, Nab and abBA form a complete program of
measurements of the main neutron decay parameters in a single
apparatus with shared systematics and consistency checks.  The
experiments are complementary: Nab is highly optimized for the
measurement of $a$ and $b$, while abBA focuses on $A$ and $B$ with a
lower-precision consistency check of the $a$ and $b$ parameters.  Nab
joins two existing experiments, aSPECT \cite{Bae08} and aCORN
\cite{Wie05}, which also study
 $a$.

\section{Measurement Principles and  Apparatus}
\label{sec:meas_princ_app}

The correlation parameter of interest, $a$, measures the dependence of
the neutron beta decay rate on the cosine of the e--$\nu$ relative
angle.  The Nab method of determination of $a$ relies on the linear
dependence of $\cos\theta_{\rm e \nu}$ on $p_{\rm p}^2$, the square of
the proton momentum for a given electron momentum (or energy).  
Conservation of momentum gives the relation
\begin{equation}
  p_{\rm p}^2
   = p_{\rm e}^2 + 2p_{\rm e}p_\nu\cos\theta_{\rm e\nu} + p_\nu^2 \,,
    \label{eq:psq_cos_enu}
\end{equation}
\begin{figure}[b]
 \noindent\hbox to \columnwidth{\hfill
   \includegraphics[width=0.9\columnwidth]{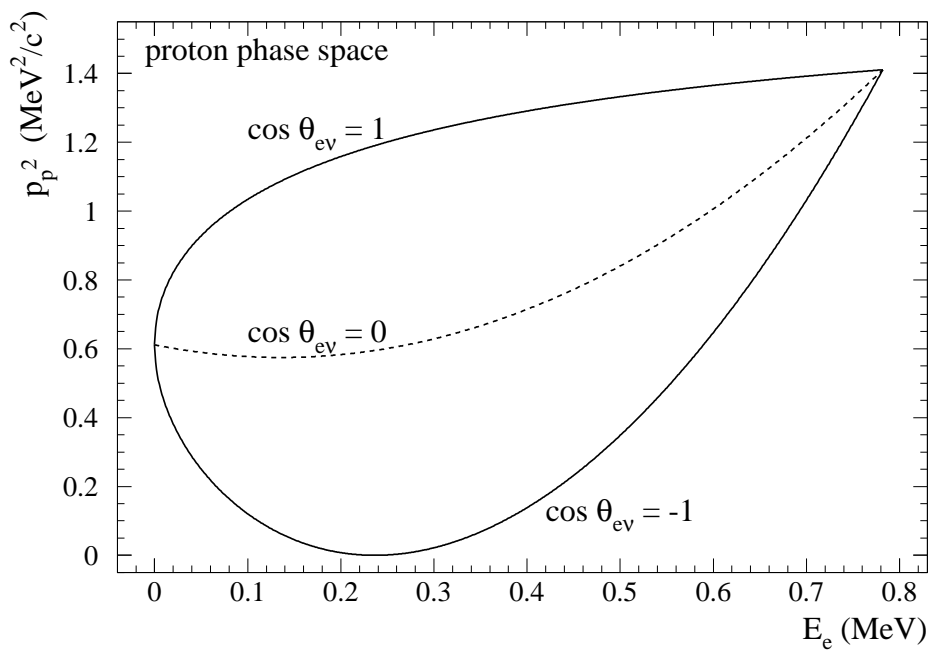}
                    \hfill}
 \caption{Proton phase space (in terms of $p_{\rm p}^2$) in neutron
 beta decay as a function of electron kinetic energy.  The upper bound
 of the allowed phase space occurs for collinear e and $\nu$ momenta,
 $\cos\theta_{\rm e\nu}=1$, while the momenta are anticollinear,
 $\cos\theta_{\rm e\nu}=-1$, at the lower bound.  The central dashed
 parabola corresponds to orthogonal $e$ and $\nu$ momenta. 
 }
 \label{fig:proton_PS}
\end{figure}
\begin{figure}[th]
 \noindent\hbox to \columnwidth{\hfill
   \includegraphics[width=0.9\columnwidth]{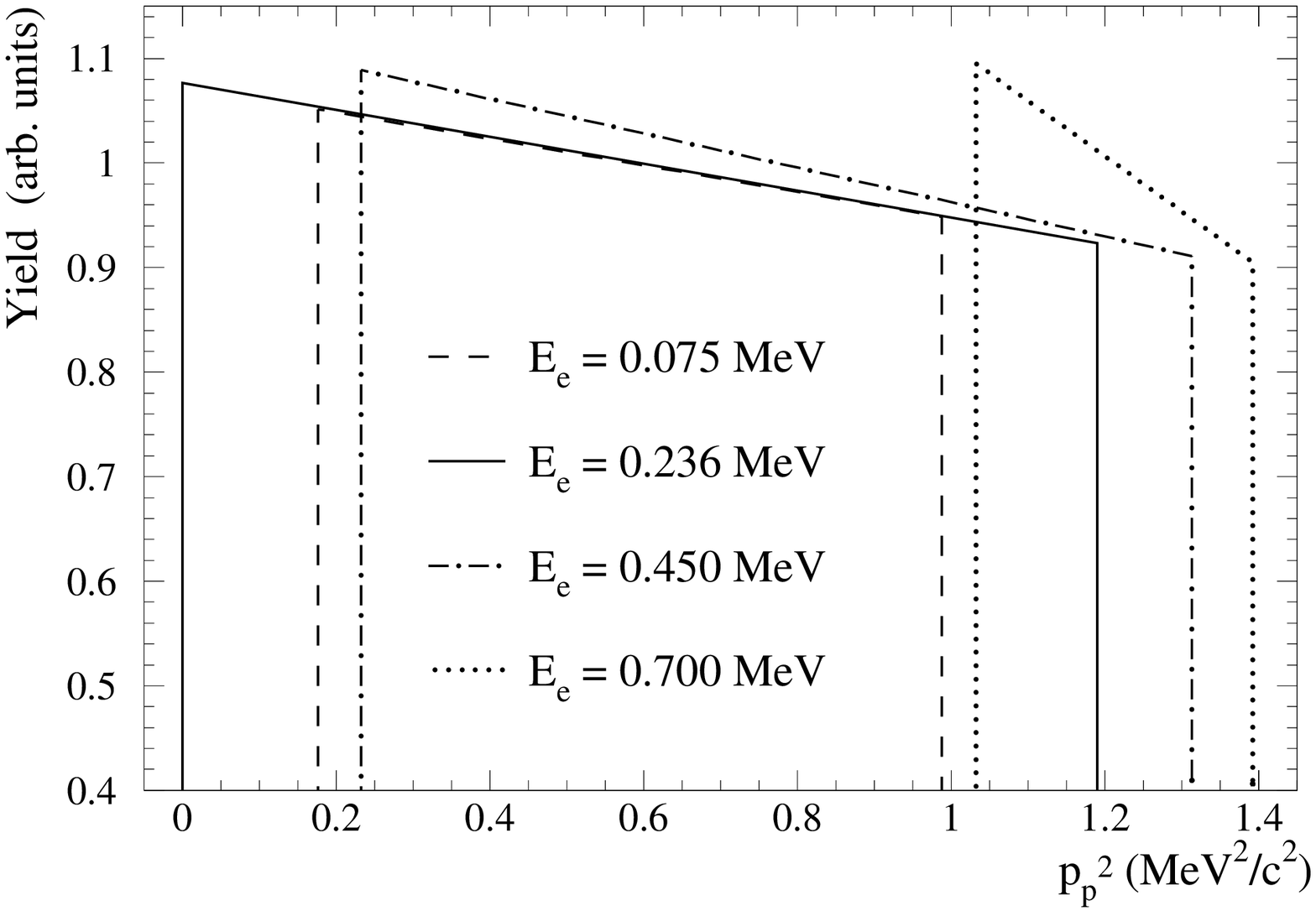}
                    \hfill}
 \caption{A plot of proton yield for four different electron kinetic
 energies with $a = -0.105$.
 If $a$ were 0, all the  distributions would have a slope of 0.
 Vertical scale origin is suppressed.} 
 \label{fig:resp_slope}
\end{figure}
where, to a very good approximation, $p_\nu$ depends only on $E_{\rm
e}$ (or $p_{\rm e}$).  Thus, Eq.~\ref{eq:psq_cos_enu} reduces to a
linear relation between $\cos\theta_{\rm e\nu}$ and $p_{\rm p}^2$ for
a fixed $p_{\rm e}$.  The mapping of $\cos\theta_{\rm e\nu}$ and
$p_{\rm p}^2$ is shown graphically in Fig.~\ref{fig:proton_PS}.  In
this plot, the phase space alone distributes proton events evenly in
$p_{\rm p}^2$ between the lower and upper bounds for any fixed value
of $E_e$.  Given the linear relationship between $p_{\rm p}^2$ and
$\cos\theta_{\rm e\nu}$, the slope of the $p_{\rm p}^2$ probability
distribution is determined by the correlation parameter $a$; in fact
it is given by $\beta a$, where $\beta = v_{\rm e}/c$ (see
Fig.~\ref{fig:resp_slope}).  This observation leads to the main
principle of measurement of $a$ which involves measurement of the
proton momenta via the proton time of flight (TOF), $t_{\rm p}$, in a
suitably constructed magnetic spectrometer.  Ideally, the magnetic
field longitudinalizes the proton momentum and $t_{\rm p} \propto
1/p_{\rm p}$; $t_{\rm p}$ is measured as the difference between the
arrival times of the electron and the proton at the detector(s).  In
the present discussion we neglect the electron TOF.  Parameter $a$ is
determined from the slopes of the $1/t_{\rm p}^2$ distributions for
different values of $E_{\rm e}$.  If $a$ were null, all distributions
would have a slope of zero.  Having multiple independent measurements
of $a$ for different electron energies provides a powerful check of
systematics, as discussed below.  The Fierz interference term $b$ is
determined from the shape of the measured electron energy spectrum.

For fixed $E_{\text{e}}$, a perfect spectrometer would record a
trapezoidal distribution of $1/t_{\rm p}^2$ with sharp edges.  The
precise location of these edges is determined by well-defined
kinematic cutoffs that only depend on $E_{\text{e}}$.  However, a
realistic time-of-flight spectrometer will produce imperfect
measurements of the proton momenta due to the spectrometer response
function, discussed in Sect.~\ref{sec:uncerts}.  The measured
locations and shapes of edges in $1/t_{\rm p}^2$ distributions will
allow us to examine the spectrometer response function and verify that
the fields have been measured correctly.

The main requirements on the spectrometer are:

\begin{enumerate}

\item The spectrometer and its magnetic ($\vec{B}$) and electric
  ($\vec{E}$) fields are designed to be azimuthally symmetric about
  the central axis, $z$.

\item Neutrons must decay in a region of large $\vec{B}$.  The
  resulting protons and electrons spiral around a magnetic field line.

\item An electric field is required to accelerate the proton from the
  eV-range energies to a detectable energy range prior to reaching the
  detector.  This field imposes, however, an energy threshold on e$^-$
  detection.

\item The proton momentum must rapidly become parallel to the magnetic
  field direction to ensure that the proton time of flight $t_{\rm p}
  \propto 1/p_{\rm p}$.  This requirement dictates a sharp field
  curvature ($d^2B_z/dz^2$) at the origin, followed by a sharp falloff
  of $B_z$.

\end{enumerate}

The basic concept of the spectrometer consists of collinear solenoids
with their longitudinal axis oriented normal to the neutron beam,
which passes through the solenoid center.  The solenoidal magnetic
field starts out high at the position of the neutron beam, typically
4\,T, dropping off quickly to parallelize the momenta as protons enter
the long ``drift'' region.  In the detection region at either end of
the solenoid the field is increased to 1/4 of its central peak value.
Cylindrical electrodes (consisting of three sections) maintain the
neutron decay region at a potential of +30\,kV with respect to the
ends of the solenoid where detectors are placed at ground potential.

The magnetic field strength is sufficiently high to constrain both
electrons and protons from neutron decay to spiral along the magnetic
field lines with the component of the spiral motion transverse to the
field limited by cyclotron radii of the order of a few millimeters.

\begin{figure}[t]
\centering \includegraphics[width=0.9\columnwidth]{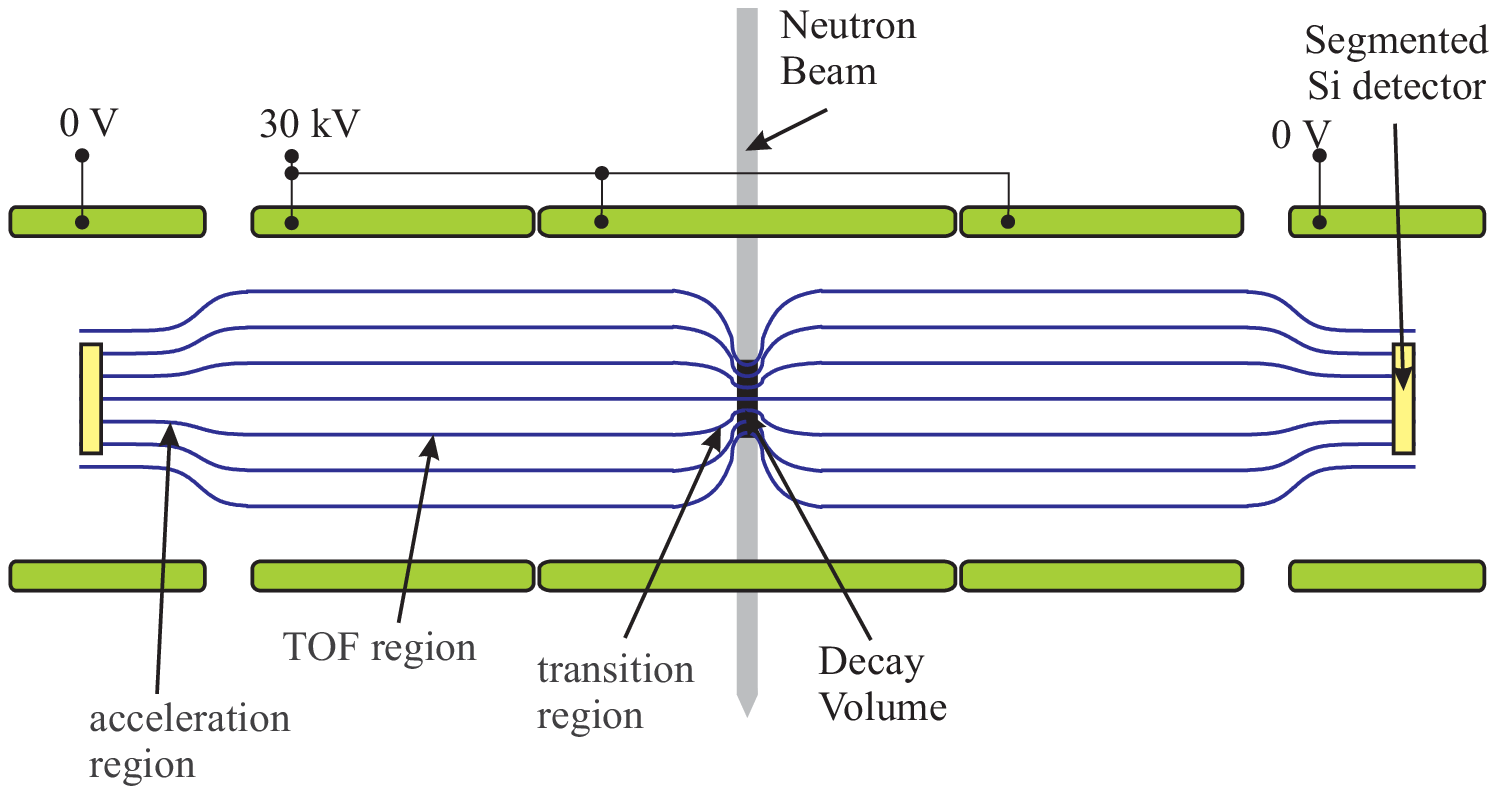}
\centering \includegraphics[width=0.9\columnwidth]{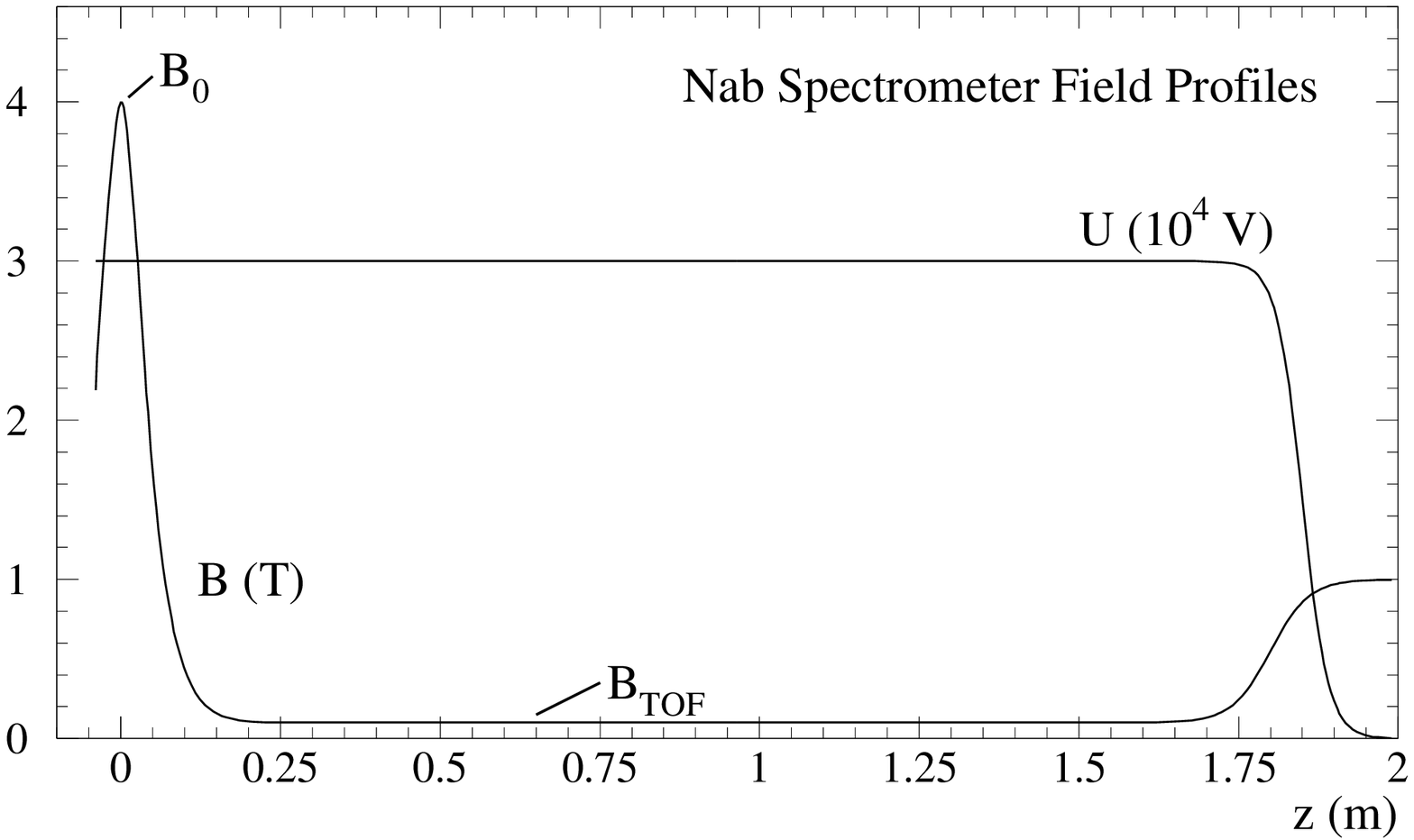}
\caption{Top panel: A schematic view of the vertical field expansion
  spectrometer showing the main regions of the device: (a) neutron
  decay region, (b) transition region with expanding magnetic field,
  (c) drift (TOF) region, and (d) the acceleration region before the
  detector.  Bottom panel: Electrical potential ($U$) and magnetic
  field ($B$) profiles on axis for 1/2 of the Nab spectrometer
  length.} 
   \label{fig:spect_1}
\end{figure}
Hence, two segmented Si detectors, one at each end of the solenoid,
view both electrons and protons in an effective $4\pi$ geometry.  The
time of flight between the electron and proton is accurately measured
in a long, $l \sim\,1.5$ meter, drift distance.  The electron energy
is accurately measured in the Si detectors.  The proton momentum and
electron energy determine the electron--neutrino opening angle.  We
note that by sorting the data on proton time of flight and electron
energy, $a$ can be determined with a statistical uncertainty that is
only 4\,\% greater than the theoretical minimum \cite{nab_prop}.

\begin{figure}[b]
 \begin{center}
  \includegraphics[width=0.7\columnwidth]{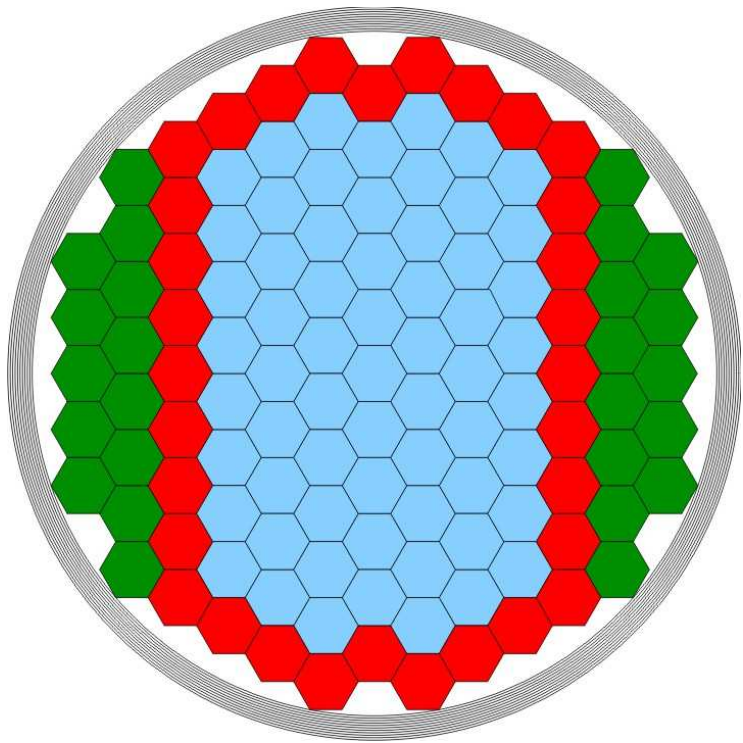}
 \end{center}
  \caption{Design for the ohmic side of the detector. The 127 hexagons
    represent individual detector elements. Proton events in the
    interior hexagons generate a valid trigger, while the perimeter
    hexagons are used only for detecting electrons. The concentric
    circles represent the guard ring structure. Electrical contact is
    made to each hexagon to provide the bias voltage and collect the
    charge deposited by incident particles. The areas between the
    pixels and guard rings are electrically connected to form one
    additional channel.}
 \label{fig:det_sketch}
\end{figure}
\begin{figure}[t]
\centering \includegraphics[width=0.7\columnwidth]{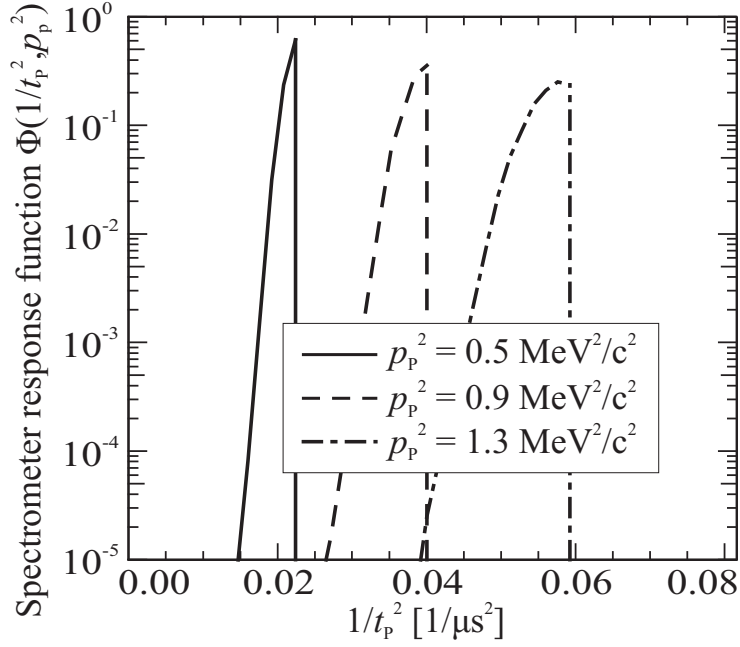}
\caption{Nab spectrometer response function $\Phi$, shown for
    different proton momenta, the magnetic field from
    Fig.~\ref{fig:spect_1} and a centered neutron beam with a width of
    2\,cm.  The calculation assumes full adiabaticity of the proton
    motion.}
\label{fig:NabSpectrometerResponse}
\end{figure}
A not-to-scale schematic view of the field expansion spectrometer is
shown in Fig.~\ref{fig:spect_1}.  Electrons and protons spiral around
magnetic field lines and are guided to two segmented Si detectors,
each having a $\sim$100\,cm$^2$ active area, and depicted
schematically in Fig.~\ref{fig:det_sketch}.  In the center of the
spectrometer the axial field strength is 4\,T, in the drift region
0.1\,T, and near the Si detectors 1\,T (see Fig.~\ref{fig:spect_1}).

\begin{figure}[h]
\centering \includegraphics[width=0.85\columnwidth]{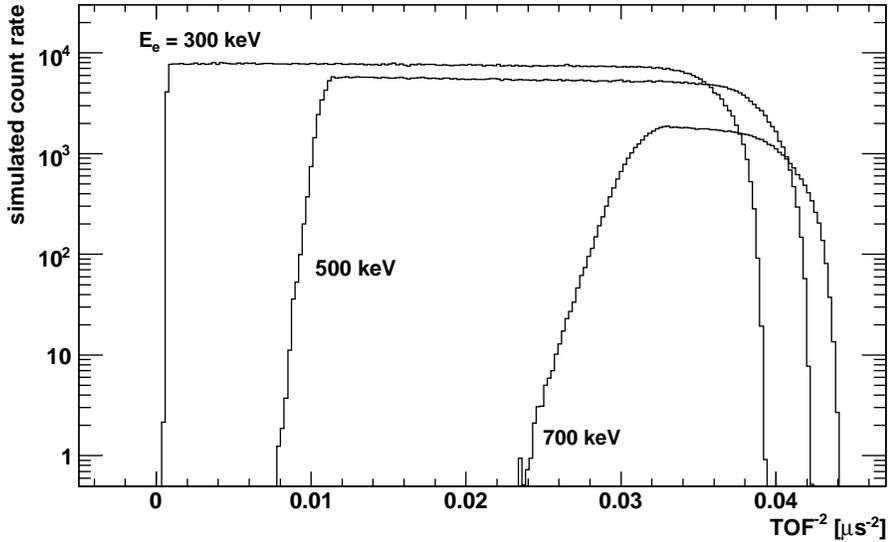}
\caption{Proton TOF spectra, $P_t(1/t_{\rm p}^2)$, for electron
    kinetic energies $E_{\rm e} = 300$, 500 and 700\,keV, generated in
    a realistic GEANT4 Monte-Carlo simulation using the $\vec{B}$
    field from Fig.~\ref{fig:spect_1} and a centered neutron beam with
    a width of 2\,cm.}
\label{fig:RealSpec}
\end{figure}

In a realistic spectrometer, however, the perfect one-to-one
correspondence of proton momentum and time of flight is lost, due to
imperfect momentum longitudinalization and other systematic effects,
such as the lateral size of the neutron beam in the decay region.  In
other words, the detector response function instead of being a delta
function in $1/t_{\rm p}^2$ for each value of $p_{\rm p}^2$, becomes a
broadened function, such as the ones calculated for three proton
momenta and depicted in Fig.~\ref{fig:NabSpectrometerResponse}.
The key challenge of the Nab approach to measuring $a$ is to minimize
the width of the detector response function while keeping the relevant
systematics under control.  The resulting TOF distributions no longer
have sharply cut off edges as in Fig.~\ref{fig:resp_slope}.  A sample
set of results of GEANT4~\cite{GEANT4} Monte Carlo calculations for
three electron energies is shown in Fig.~\ref{fig:RealSpec}.

Strictly speaking, determining $b$ requires detecting only the
electron and reliably measuring its kinetic energy.  Nevertheless,
there are a number of challenges associated with this measurement,
commented on in the following Section.

\section{Measurement Uncertainties and Systematics}
\label{sec:uncerts}

The statistical sensitivity of our measurement method is primarily
determined by the spectrometer acceptance and imposed energy and TOF
restrictions.  The statistical uncertainties for our measurements of
the $a$ and $b$ parameters in neutron decay are listed in
Tab.~\ref{tab:stat_unc}, reflecting the dependence on $E_{\rm e,min}$,
the electron kinetic energy detection threshold, and $t_{\rm p,max}$,
the maximum proton TOF accepted.  Additionally, the electron energy
calibration $E_{\rm cal}$ and the precise length $l$ of the low-field
drift region represent important sources of systematic uncertainty.
Thus, parallel analyses will be performed keeping $E_{\rm cal}$ and
$l$ free, in order to study and remove their systematic effects.
Table~\ref{tab:stat_unc} shows that the reduction in statistical
sensitivities under these conditions is modest.

\begin{table}[b]
\caption{Top: statistical uncertainties $\sigma_a$ for the e-$\nu$
  correlation parameter $a$.  A perfect spectrometer would obtain
  $\sigma_a = 2.3/\sqrt{N}$.  Bottom: statistical uncertainties
  $\sigma_b$ for the Fierz interference term $b$.}
\setlength{\adlt}{0.092\columnwidth}  
  \begin{tabular}{l@{\extracolsep{\adlt}}c@{\extracolsep{\adlt}}c@{\extracolsep{\adlt}}c@{\extracolsep{\adlt}}c}
    \hline
    $E_{\rm e,min}$  & 0 & 100\,keV &  100\,keV & 300\,keV \\
    $t_{\rm p,max}$  & -- &   --  &  $10\,\mu$s & $10\,\mu$s \\
    \hline\hline 
    $\sigma_a$
    & $2.4/\sqrt{N}$ & $2.5/\sqrt{N}$ & $2.6/\sqrt{N}$ & $3.5/\sqrt{N}$ \\
    $\sigma_a^\dagger$ & $2.5/\sqrt{N}$ & $2.6/\sqrt{N}$ & -- & -- \\[0.5ex]
    \hline
    \hline
    $E_{\rm e,min}$  & 0 & 100\,keV &  200\,keV & 300\,keV \\
    \hline\hline 
    $\sigma_b$ & $7.5/\sqrt{N}$ & $10.1/\sqrt{N}$ & $15.6/\sqrt{N}$
                                                  & $26.4/\sqrt{N}$ \\
    $ \sigma_b^{\dagger\dagger}$
             & $9.1/\sqrt{N}$ & $12.7/\sqrt{N}$
                             & $20.3/\sqrt{N}$ & $35.1/\sqrt{N}$ \\[0.5ex]
    \hline
    \end{tabular} \\
    $^\dagger$ with $E_{\rm cal}$ and $l$ variable. \quad 
    $^{\dagger\dagger}$ with $E_{\rm cal}$ and $a$ variable.
\label{tab:stat_unc}

\end{table}

The calculated FnPB neutron decay rate under SNS full-power conditions
of $\sim$19.5/(cm$^3$s), and with the Nab fiducial decay volume of
20\,cm$^3$, yields $\sim$400 detected decays/sec \cite{Huf05}.  In a
typical 10-day run of $7\times 10^5$\,s of net beam time we would
achieve $\sigma_a/a \simeq 2 \times 10^{-3}$ and $\sigma_b \simeq 6
\times 10^{-4}$.  Since we plan to collect several samples of $10^9$
events in several 6-week runs, the overall Nab accuracy will not be
statistics-limited.

Controlling the measurement systematics presents by far the greatest
challenge in the Nab experiment.  The most basic task is to specify
the spectrometer fields with precision sufficient for an accurate
determination of the spectrometer response function $\Phi(1/t_{\rm
p}^2,p_{\rm p}^2)$.  We have adopted two methods of addressing this
problem.  In the first approach (Method A), we determine the shape of
the spectrometer response function from theory, leaving several
parameters free, to be determined by fits to the measured spectra.
The second approach (Method B) relies on obtaining the detection
function with its uncertainties {\slshape a priori} from a full
description of the neutron beam and electromagnetic field geometry.
Subsequently, the experimental data are fitted with only the physics
observables as free parameters.  Below we summarize some of the main
challenges along with strategies for their control at the required
level.  A much more detailed discussion of both methods and the
experimental challenges is given in the Nab experiment
proposal~\cite{nab_prop}.


\smallskip\noindent 
\underline{Uncertainties in $a$ due to the spectrometer response}

  \begin{itemize}
   \item {\em Precise specification of the neutron beam profile:} A
     mere 100\,$\mu$m shift of the beam center induces $\Delta a/a
     \sim 0.2\,$\%.  However, this effect cancels when averaging over
     the two detectors on opposite sides of the solenoid; measuring a
     nonzero up-down proton counting asymmetry pins it down
     sufficiently.

   \item {\em Magnetic field map:} The field expansion ratio defined
     as $r_\text{B}=B_\text{TOF}/B_0$ must be controlled at the level
     of $\Delta r_\text{B}/r_\text{B} =10^{-3}$ in order to keep
     $\Delta a/a$ under $10^{-3}$.  This will be mapped out using a
     calibrated Hall probe.  Field curvature must be determined with
     an accuracy of $1\times 10^{-3}$ in dedicated measurements.
     Average mapping accuracy $\Delta B/B$ must be kept below $\sim 2
     \times 10^{-3}$.

   \item {\em Flight path length:} An uncertainty of order $\Delta l \leq
     30\,\mu$m results in $\Delta a/a$ at our limit.  Hence, $l$ will
     be kept as a fitting parameter.  Additionally, we will perform a
     consistency check by making differential measurement using
     segmented electrodes.

   \item {\em Homogeneity of the electric field:}  Electric potential
     will have satisfy stringent limits on inhomogeneities as
     discussed in the Nab proposal \cite{nab_prop}.

   \item {\em Rest gas:} requires vacuum of $10^{-7}$\,Pa or better.

   \item {\em Adiabaticity} of the magnetic field configuration is not
     an absolute requirement.  Detailed Monte Carlo analysis has shown
     excellent efficiency of proton momentum longitudinalization for
     certain relatively non-adiabatic fields.  However, an adiabatic
     design makes the evaluation of systematic errors simpler and more
     reliable.

  \item {\em Doppler effect:} Adverse effects of the Doppler effect
    will apparently be controlled sufficiently by the spectrometer
    design, but a thorough analysis will be made in conjunction with
    the final design.
  \end{itemize}
 
\smallskip\noindent
\underline{Uncertainties in $a$ due to the detector}

  \begin{itemize}
   \item {\em Detector alignment:} The spectrometer imaging properties
     provide for a self-consistent calibration in the data.
   \item {\em Electron energy calibration} is required at the
     $10^{-4}$ level.  To achieve it we'll use radioactive sources,
     evaluate directional count rate asymmetries, and also leave it as
     a fitting parameter with acceptably small loss of statistical
     sensitivity (see Tab.~\ref{tab:stat_unc}).
   \item {\em Trigger hermiticity} is affected by the particle impact
     angle on the detector, backscattering, and TOF cutoff (planned in
     order to reduce accidental backgrounds).  Several consistency
     checks will be evaluated from the data to quantify and
     characterize the various aspects of trigger hermiticity.
   \item {\em TOF measurement uncertainties:} The requirement is
     $\Delta (t_{\rm p}-t_{\rm e}) \sim 100$\,ps.  While it is not
     necessary to reach this timing accuracy for each event, it has to
     be achieved for the event sample average, a realistic goal given
     the planned event statistics.
   \item {\em Edge effects} introduce important systematics.  Thanks
     to the imaging properties of the spectrometers, these can be
     controlled and corrected for to a sufficient degree with
     appropriate cuts on the data.
  \end{itemize}

\smallskip\noindent
\underline{Uncertainties in $b$} \\
     Sources of uncertainties in the measurement of $b$ are fewer than
     for $a$ since accurate proton momentum measurement (via its TOF)
     is not required.
     The dominant sources are electron energy calibration (discussed
     above) and electron backgrounds.

\smallskip\noindent
\underline{Backgrounds for $a$ and $b$}

  \begin{itemize}
   \item {\em Neutron beam related backgrounds} are notoriously hard
     to calculate and model {\slshape a priori}, and will ultimately
     have to be measured and characterized {\slshape in situ}.
     Reasonable estimates place the beam-related background rates
     below the signal rate.  While we have plans for shielding and
     lining surfaces with neutron absorbing $^6$LiF material, the
     coincident technique of detecting e--p pairs helps to reduce
     substantially the effect of beam-related accidental backgrounds.
   \item {\em Particle trapping:} Electrons can be trapped in the
     decay volume, expansion, and TOF regions.  These regions form an
     electrode-less Penning trap.  The potential well trap does not
     cause a problem for electrons above our energy threshold.  The
     longitudinalization of the electron momentum due to the magnetic
     field allows all of them to escape and to reach the detector.
     Low energy electrons from neutron decay, from field ionization or
     from rest gas interactions are a concern since trapped particles
     ionize the rest gas, and the ions form a time-dependent
     background.  Several strategies are under consideration to remove
     the trapped particles; they will be refined under real running
     conditions.
  \end{itemize}


\section{Summary}

The Nab collaboration plans simultaneous high-statistics measurements
of neutron decay parameters $a$, the e--$\nu$ correlation coefficient,
and $b$, the Fierz interference term, with $\Delta a/a \simeq 10^{-3}$
and $\Delta b \simeq 3\times 10^{-3}$.

Basic properties of the Nab spectrometer are well understood; details
of the fields are under study in extensive analytical and Monte Carlo
calculations.  

Elements of the proposed Nab spectrometer will be shared with other
neutron decay experiments, such as abBA.

Development of the abBA/Nab Si detectors is ongoing and remains a
technological challenge.  Each of the target properties of the
detector have been realized separately; the remaining task is to
realize them simultaneously in one piece of silicon.

The major elements of the data acquisition system have been
successfully developed.

The experiment received approval in Feb.\ 2008.  Under the most
favorable funding and technical scenario it could be ready for
commissioning in 2010.



\end{document}